\documentclass[twocolumn,showpacs]{revtex4}
\usepackage[xdvi]{graphicx}
\usepackage{dcolumn}
\usepackage{amsmath}

\newcommand{\bra}{\left\langle}
\newcommand{\ket}{\right\rangle}
\newcommand{\pder}[2]{\frac{\partial #1}{\partial  #2}}

\newcommand{\der}[2]{\frac{d #1}{d  #2}}
\newcommand{\bv}[1]{{\boldsymbol #1}}
\newcommand{\tauD}{\tau_{\rm D}}
\begin{document}

\title{
A fluctuation-response relation of many Brownian particles 
under non-equilibrium conditions}
\author{Takenobu Nakamura}
\affiliation
{ Research Institute of Computational Sciences ( RICS )
National Institute of Advanced Industrial Science and Technology ( AIST )
1-1-1 Umezono, Tsukuba 305-8568
JAPAN}
\author{Shin-ichi Sasa}
\affiliation
{Department of Pure and Applied Sciences,  
University of Tokyo, Komaba, Tokyo 153-8902, Japan}

\date{\today}

\begin{abstract} 
We study many interacting Brownian particles 
under a tilted periodic potential.
We numerically measure the linear response coefficient 
of the density field by applying a slowly varying 
potential transversal to the tilted direction.
In equilibrium cases, the linear response coefficient 
is related to the intensity of density fluctuations 
in a universal manner, 
which is called a fluctuation-response relation.
We then report numerical evidence that 
this relation holds 
even in non-equilibrium cases.
This result suggests 
that Einstein's formula on density fluctuations 
can be extended to driven diffusive systems
when the slowly varying potential is applied
 in a direction transversal to the driving force.
\end{abstract}
\maketitle

\section{introduction}

The construction of statistical mechanics applicable to
non-equilibrium systems has been attempted for a long time.
However, even in non-equilibrium steady states, no general 
framework is known, except for the linear response theory \cite{Kubo}.
For example, a useful form of the steady state distribution 
function has not been determined in systems far from equilibrium.

In order to consider a strategy for investigating the statistical mechanics
of non-equilibrium steady states, let us recall the justification 
of equilibrium statistical mechanics. The equilibrium statistical 
mechanics leads to Einstein's formula on macroscopic fluctuations,
which  relates a large deviation
functional of macroscopic fluctuations with a thermodynamic
function. Since the validity of this formula 
can be checked by measuring the fluctuations of macroscopic 
quantities, we expect that its examination 
in non-equilibrium
steady states might provide a hint to develop non-equilibrium
statistical mechanics.

With regard to the large deviation functionals in 
non-equilibrium steady states, we review a few  works.
First, a large deviation functional of density fluctuations
is exactly derived for non-equilibrium lattice gases
\cite{DLS1,DLS2}. This
functional is nonlocal in contrast to equilibrium cases,
and the local part takes the same form as the equilibrium form.
The latter property seems to be specific to the model, which is 
sufficiently simple to solve. Indeed, the local part of the density 
fluctuations  in a driven lattice gas is strongly influenced 
by an externally driven force \cite{HS1}. 
However, even in this case, 
it has been numerically shown that the local part can be
described by an extended, operationally constructed 
thermodynamic function \cite{HS1}.
This result has been proved mathematically for 
some systems under a special condition \cite{SST}.

In this paper, in order to check the generality of the result
in Refs. \cite{HS1} and \cite{SST}, we study a model of
interacting Brownian particles under non-equilibrium 
conditions. Here, note that the motion of Brownian 
particles is believed to be described by Langevin equations. 
This has been confirmed experimentally by measuring
the trajectories of Brownian particles with a high accuracy,
which might have become possible by the recent development 
of technology for  optical instruments \cite{KTG,Grier,Grier2}.
From this fact, the system of Brownian particles is regarded as
an ideal system for studying  the fundamental problems of 
statistical mechanics, both theoretically and experimentally.

This paper is organized as follows.
In Sec. \ref{model:num}, we introduce Langevin equations
describing the motion of many Brownian particles 
under an external driving force. We also define the correlation coefficient
$C$ and the response coefficient $R$ that we study. After reviewing
briefly, the equality between $C$ and $R$ (called a 
fluctuation-response relation), we conjecture in  Sec. \ref{conjecture:num} 
that the fluctuation response relation can be extended to
non-equilibrium systems on the basis of the consideration 
of Einstein's formula on macroscopic fluctuations.
In Sec. \ref{sokutei}, we report a result of numerical experiments.
The final section is devoted to a few remarks.

\section{model}\label{model:num}

We study the system that consists of $N$ Brownian particles suspended 
in a two dimensional solvent of temperature $T$.
Let $\bv x_i$, $i=1,2,...,N$,  be the position 
of the $i$-th particle, where  $\bv x_i\in[0,L]\times[0,L]$
with a periodic boundary conditions. 
We express the $\alpha$-th component of $\bv x_i$ 
as $x_{i\alpha}$ with $\alpha=1,2$. 
That is, $\bv x_i=(x_{i1},x_{i2})$. 
Each particle is driven by an external force
$f\bv e_1=(f,0)$ and is subject to a periodic potential
$U(x_1)$ with period $\ell$.
For simplicity, we assume that the periodic
potential is independent of $x_2$. 
Furthermore, we express the
interaction between the $i$-th and $j$-th particles by an
interaction potential $u(|\bv x_i-\bv x_j|)$.

The motion of the $i$-th Brownian particle is assumed to be
described by a Langevin equation 
\begin{align}
\gamma\der{x_{i \alpha}}{t}=&
\left(f-\pder{U(x_{i1})}{x_{i \alpha}}\right)\delta_{\alpha 1}\nonumber\\
&-\pder{ \ }{x_{i \alpha}}
\sum_{j=1,j\neq i}^{N}u(|\bv{x}_i-\bv{x}_j|)
+R_{i \alpha}(t),
\label{lange1}
\end{align}
where $\gamma$ is a friction constant and $R_{i\alpha}(t)$
is zero-mean Gaussian white noise that satisfies 
\begin{align}
\bra R_{i\alpha}(t)R_{j\beta}(t')\ket=2\gamma T\delta_{ij}\delta(t-t').
\end{align}
Here, the Boltzmann constant is set to unity.
The $i$-th and $j$-th particles interact
via the soft core repulsive potential $u(r)$ given by
\begin{align}
u(r)=\begin{cases}
K(r-2r_c)^2/2&
\text{when \ \ $r\le 2r_{\rm c}$},\\
0&\text{when \ \ $r > 2r_{\rm c}$}.
\end{cases}
\label{interaction:simulation}
\end{align}
where $r_c$ is the cutoff radius of the potential.
We also assume 
a form of the periodic potential $U(x_1)$ as
\begin{align}
U(x_1)=U_0\sin\left(\frac{2\pi x_1}{\ell}\right).
\end{align}

It should be noted that without the periodic potential $U(x_1)$,
the system is equivalent to an equilibrium system 
in a moving frame with velocity $f/\gamma$. 
Thus, the periodic potential is
necessary for investigating the non-equilibrium nature. 
In the case that $f=0$, the Langevin equation given in Eq. (\ref{lange1})
satisfies the detailed balance condition with respect
to the canonical distribution; consequently, the 
stationary distribution is canonical. On the contrary, 
in the non-equilibrium cases where $f\neq 0$,  
the stationary distribution is not the canonical distribution
because  of the lack of the detailed balance condition.

In this paper, all quantities are converted to dimensionless 
forms by setting $\gamma$, $\ell$, and $T$ to unity.
The values of the parameters in our model
are chosen as follows.
First, $r_c=2.0$, that is,  the interaction range of the particle 
is comparable in size to the period of the periodic potential.
Second, in the soft-core repulsive interaction $u(r)$ 
in Eq. (\ref{interaction:simulation}), 
$K=200,300,400$, and $500$.
Third, $U_0=15.0$ and $20.0$, and finally $L=30$ and $N=50$.
Furthermore, Eq. \eqref{lange1} is solved numerically 
by using an explicit time integration method 
with the time step $\Delta t=1.0\times 10^{-3}$.

\begin{figure}
\begin{center}
\includegraphics[width=1.0\hsize]{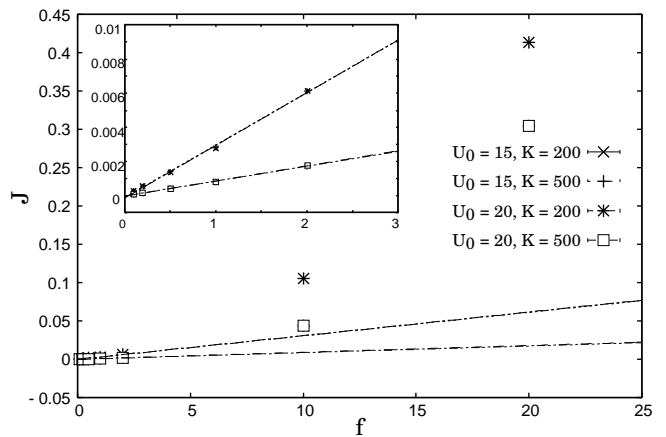}
\end{center}
\caption{Current $J$ in the $x_1$ direction 
as a function of $f$. The average values of  $J$ 
for 10 samples are plotted for several values of $f$.
The statistical error bars are smaller than the 
symbols. Inset: Close-up of the linear response regime.}
\label{current15}
\end{figure}

With these parameters, we first perform preliminary measurements
so as to check how far the system is from the equilibrium. 
Concretely, we measure  the current $J$ as a function of an external 
force $f$, where we calculate $J$ numerically as 
\begin{align}
J= \frac{1}{\tau_{\rm c}} \int_{t_{\rm in}}^{t_{\rm in}+\tau_{\rm c}} dt
\sum_{i=1}^N \frac{dx_{i1}(t)}{dt}. 
\end{align}
Here, we choose the values of $t_{\rm in}$ and $\tau_{\rm c}$ so that 
the right-hand side becomes independent of these values
within the numerical accuracy we imposed. 
The result is displayed in  Fig. \ref{current15},
by which we judge that the system with $f=20$ is 
in a state far from the linear-response regime.
Thus, in the argument below, we investigate the system
with $f=20$ as an example of non-equilibrium systems. 

Now, we study  the statistical  properties of 
the density field 
\begin{align}
\rho(\bv x,t)=\sum_{i=1}^N\delta(\bv x-\bv x_i(t)).
\label{dens:num}
\end{align}
We consider the Fourier series of the density field
\begin{align}
\hat \rho_{(n_1,n_2)}=& \int d^2\bv x \rho(\bv x) 
e^{-i\frac{2\pi}{L}(n_1 x_1+n_2 x_2)}.
\end{align}
Using this,  we  characterize density fluctuations 
on large scales in terms of  the correlation coefficient
\begin{align}
C\equiv\bra\left({\rm Re}[\delta\hat\rho_{(0,1)}]\right)^2\ket,
\label{def:C}
\end{align}
where $\delta \hat\rho_{(n_1,n_2)}$ is a Fourier series of 
the deviation 
$\delta\rho(\bv x)= \rho(\bv x)-\bra \rho(\bv{x})\ket$ 
from the average density profile $\bra \rho(\bv{x})\ket$, 
and ${\rm Re}[\ ]$ represents the real part of a complex number.

According to the fluctuation-response relation, which 
is valid in equilibrium cases, $C$
is related to a response coefficient against a 
potential perturbation. In the present problem, 
we consider the response of 
${\rm Re}[\delta\hat\rho_{(0,1)}]$ to the potential 
\begin{align}
V(\bv x)=\epsilon \cos\left(\frac{2\pi x_2}{L}\right)
\label{V2}.
\end{align}
We note that this potential does not depend on $x_1$,
that is, we apply the perturbation potential transversal 
to the external field $f$ in a manner similar to that in
Refs. \cite{HS1,SST}. Quantitatively, the response
coefficient is defined as 
\begin{align}
R\equiv -\lim_{\epsilon\rightarrow 0}
\frac{\bra{\rm Re}[\delta \hat\rho_{(0,1)}]\ket_\epsilon}{\epsilon},
\label{def:R}
\end{align}
where $\bra \  \ket_\epsilon$ represents the statistical average 
in the steady state under  the potential (\ref{V2}).

Here, by the fluctuation-response relation is meant 
\begin{align}
C=TR\label{FRR}.
\end{align}
Although it should hold in equilibrium cases, 
there is no reason to expect that it also holds 
in non-equilibrium systems. Nevertheless, we can 
measure both $C$ and $R$ even for non-equilibrium 
systems. Thus, we can check the relation 
concretely by numerical experiments. 

\section{conjecture}\label{conjecture:num}

Before presenting the results of numerical experiments,
we review the understanding of the relation given in Eq. (\ref{FRR}).
In equilibrium cases, the relation can be directly obtained from the 
canonical distribution; however, here we present an alternative, 
phenomenological understanding based on a macroscopic fluctuation theory
in order to consider an extension of the relation.

Let $\tilde\rho(\bv{x})$ be a fluctuating density field 
defined at a macroscopic scale. 
Then, due to the large deviation property,
the stationary distribution is written as
\begin{align}
P_{\rm S}(\tilde\rho(\cdot)) \simeq 
\exp\left(- L^2 I( \tilde \rho(\cdot) ) \right),
\label{LD}
\end{align}
where $I( \tilde\rho(\cdot) )$ is called a large deviation
functional. According to the fluctuation theory for systems 
under equilibrium conditions, $I$ is determined by thermodynamics.
Concretely, using a free energy density $f(T,\rho)$, we can write
\begin{align}
I(\tilde\rho(\cdot))=\frac{1}{TL^2} \int d^2\bv{x} 
&[f(T,\tilde\rho(\bv{x}))-f(T,\bar \rho)\nonumber\\
&+V(\bv x)(\tilde\rho(\bv x)-\bar \rho)],
\label{Einstein2}
\end{align}
where $\bar \rho$ is the average density. This relation is called 
Einstein's formula and can be proved 
within a framework of equilibrium statistical mechanics.
Noting that $\hat\rho_{(0,1)}$ is an example of macroscopic
density fluctuations, we find that  Eq. (\ref{FRR}) is 
obtained from Eqs. (\ref{LD}) and (\ref{Einstein2}).

Equations  (\ref{LD}) and (\ref{Einstein2}) are so simple
that they are expected to be extended to those  valid 
in non-equilibrium systems. More concretely, we conjecture 
that Eq. (\ref{Einstein2}) holds if the free energy is 
extended in a consistent manner. 

In this conjecture, the existence of the extended free energy 
is highly nontrivial. Indeed, severe restrictions are 
required to construct thermodynamics in non-equilibrium 
steady states, as discussed in Ref. \cite{SST}.  
Nevertheless, we already know that thermodynamics
can be extended in non-equilibrium states for driven 
lattice gases  when we focus on the transversal direction 
to the driving force \cite{HS1, SST}. Since the present model 
has common features with driven lattice gases,  we expect  
that the thermodynamics can be extended to our model
in a manner similar to that for driven lattice gases. 

These considerations lead us to conjecture the validity 
of Eq. (\ref{FRR}), although the statistical distribution 
is not canonical in non-equilibrium cases.
Note that Eq. (\ref{FRR}) can be checked 
independently of the question whether or not
thermodynamic functions can be extended to
non-equilibrium steady states.

\section{Measurement and result}\label{sokutei}

Before measuring $C$ and $R$, we  estimate the relaxation 
time $\tauD$ for the density field. 
This estimation is necessary for the accurate determination of 
statistical averages by numerical experiments. 
Since perturbation effects of the potential $V(\bv x)$ can be neglected, 
we estimate $\tauD$ only for the systems without this potential. 
Initially, we prepare an inhomogeneous particle configuration;
as an example, all particles are placed on the sites of
a hexagonal lattice with the  lattice interval $2r_c$.
Then, we observe a diffusion phenomenon of the density field by solving
Eq. (\ref{lange1}) numerically. 
Because we found that the diffusion in the $x_1$ 
direction is slower than that in the $x_2$ direction,
we measure only the time dependence of $\hat\rho_{(1,0)}(t)$.
Repeating this procedure $30$ times, we estimate the statistical average
of $|\hat\rho_{(1,0)}(t)|$.
We then obtain a fitting 
$|\hat \rho_{(1,0)}(t)|=|\hat\rho_{(1,0)}(0)|e^{-t/\tauD}$.
We found  that the $K$ 
dependence of $\tauD$ is small and that $\tauD$ is 
a decreasing function of $f$ when the other parameters are fixed. 
From these observations, we estimate $\tauD$ for the system 
with $U_0$, $K$, and $f$ by measuring $\tauD$ for the system 
with $f=0$, $K=200$, and $V(\bv x)=0$.

Using $\tauD$ evaluated above, we estimate a statistical average 
$\bra A\ket_\epsilon$ using a time average
\begin{equation}
\bra A \ket_\epsilon =\frac{1}{\tau} \int_{t_0}^{t_0+\tau} dt A(t),
\label{average}
\end{equation}
where we chose the values of $t_0$ and $\tau$ as those  more than 40 
times of the relaxation time $\tauD$ for the density field. 

\begin{figure}
\begin{center}
\includegraphics[width=1.0\hsize]{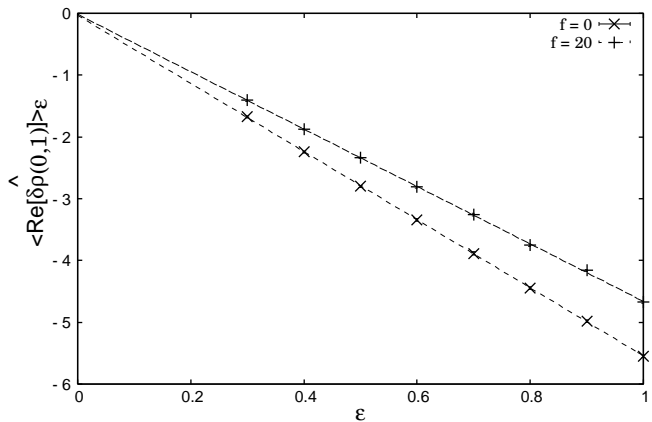}
\end{center}
\caption{
Average values of $\bra{\rm Re}[\delta\hat\rho_{(0,1)}]\ket_\epsilon$ 
over $100$ samples are plotted as a function of $\epsilon$.
$U_0=15$ and $K=200$. The statistical error bars are smaller 
than the symbols. }
\label{graph:response1024}
\end{figure}

We now measure $R$ in Eq. (\ref{def:R}) for all the values
of $U_0$ and $K$ (mentioned previously) both under the 
equilibrium condition $f=0$ and the non-equilibrium 
condition $f=20$. By measuring 
$\bra{\rm Re}[\delta\hat \rho_{(0,1)}]\ket_\epsilon$ 
for several values of $\epsilon$,
we estimate the regime where it linearly depends on $\epsilon$.
The slope in this region yields 
$R$. As one example, in Fig. \ref{graph:response1024},
$\bra {\rm Re}[\delta\hat \rho_{(0,1)}]\ket_\epsilon$ 
is displayed for $U_0=15$ and $K=200$:
$R$ is evaluated as $5.51\pm0.01$
for the non-equilibrium case ($f=20$), 
while it is $4.64\pm0.03$ for the equilibrium case.

Next, we measure 
$C=\langle({\rm Re}[\delta\hat\rho_{(0,1)}])^2\rangle_0$
for the same values of the parameters. We find that for $f=20$, 
the value of $C$ 
shows clear 
deviation from the equilibrium value. This suggests that 
the stationary distribution of $\rho$ in the non-equilibrium case is not close
to the equilibrium distribution.

\begin{figure}
\begin{center}
\includegraphics[width=1.0\hsize]{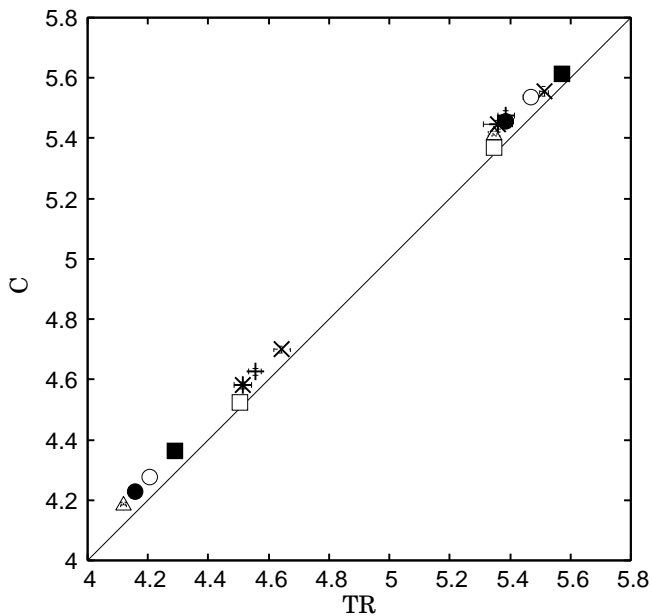}
\end{center}
\caption{$C$ vs $TR$ 
for several values of $U_0,K$, and $f$.
Two same symbols correspond to the data obtained 
by using the same $U_0$ and $K$ 
for the equilibrium and non-equilibrium conditions.}
\label{graph:FRR}
\end{figure}

From the two independent measurements of $R$ and $C$, 
we can check the fluctuation-response relation $C=TR$.
The result is summarized in Fig \ref{graph:FRR}.
This suggests that the relation given in Eq. (\ref{FRR})
holds even in the non-equilibrium steady states to  a similar extent 
as that in the equilibrium states. 

\begin{figure}
\begin{center}
\includegraphics[width=1.0\hsize]{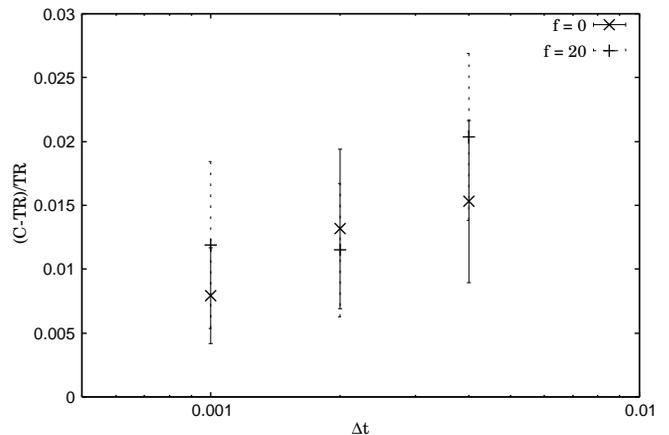}
\end{center}
\caption{$(C-TR)/TR$ as a function of the time step $\Delta t$.
$U_0=15$ and $K=200$.}
\label{graph:error}
\end{figure}

Here, we present a remark on the systematic deviation 
from the equality in Fig \ref{graph:FRR}. Let us
recall that the validity of the  equality given in Eq. (\ref{FRR})
can be proved mathematically in equilibrium cases. 
Therefore, each point in Fig. \ref{graph:FRR} should exist 
on the solid line within the statistical error bar 
at least for the equilibrium cases.
We conjecture that the slight but systematic deviation from the solid
line in Fig. \ref{graph:FRR} is caused by numerical inaccuracy 
associated with a choice of the value of $\Delta t$.
In order to check it further, we measured how the deviation depends on 
the time step $\Delta t$ in our numerical
calculation. As shown in Fig \ref{graph:error}, 
$C/(RT)$ approaches  1 in the limit 
$\Delta t \to 0$.  Thus, we expect that 
in both equilibrium and non-equilibrium cases, 
the fluctuation-response relation holds 
with a higher  accuracy if we perform numerical 
experiments with a smaller $\Delta t$.

\section{concluding remarks}\label{concludingremarks:num}

Before concluding this paper, we present two remarks. 
First, let us notice that the system we study is a typical 
example of the so-called {\it driven diffusive system}
\cite{Schmidttmann,Gregory,Garrido}.
It has been  believed that in such a system the equal-time spatial 
correlation function generally exhibits a power-law decay 
of the type $r^{-d}$ for a large distance $r$, where $d$ is the spatial 
dimension of the system \cite{Grinstein,Dorfman}.
This power-low decay is called the long-range correlation.
In fact, it has been proved for the model given in Eq. (\ref{lange1})
that this type of long-range correlation appears when the interaction 
length between two particles is sufficiently larger than the period $\ell$
\cite{NakamuraSasa}. However, such a nonlocal behavior of 
density fluctuations was not observed in our numerical experiments.
A plausible explanation for this apparent inconsistency might be
that the system size we study is much smaller than that at which the 
long-range correlation can be observed, though the system size is 
so large that thermodynamic fluctuations can be argued. Further studies
will be necessary in order to gain a clearer understanding.

\begin{figure}
\begin{center}
\includegraphics[width=1.0\hsize]{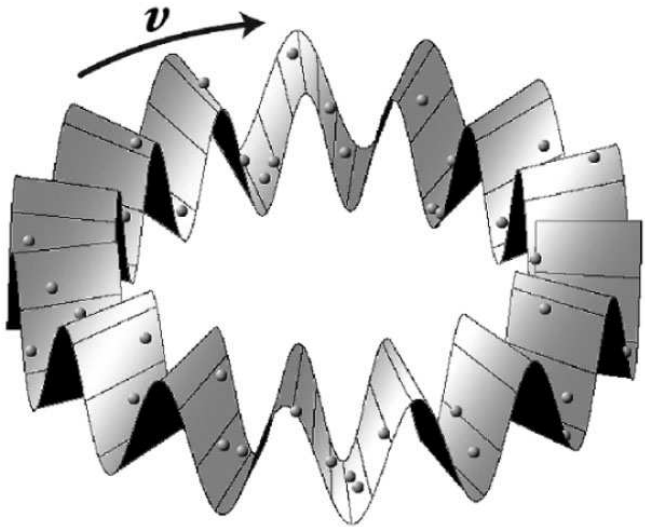}
\end{center}
\caption{
Schematic figure of a driven diffusive system under periodic 
boundary conditions.}
\label{periodic_bc}
\end{figure}

As the second remark, 
we address an example of laboratory experimental systems related
to our study and consider the possibility of observing our simulation
results experimentally.
First, since a periodic 
potential with a period of $6$ $\mu$m can be designed 
by using an optical instrument \cite{Brunner},
let us assume that $\ell=1$ corresponds to $6 \mu$m.
Then $r_c=2.0$ and $L=30$ correspond to $12.0$ $\mu$m and 
$180$ $\mu$m, respectively. 
Here,  the cutoff radius $r_c$ 
may be interpreted as an 
interaction range such as the Debye screening length of the 
screened Coulomb potential. According to Ref. \cite{Blaaderen}, 
such a long screening length can be realized experimentally.
Note that the core radius of the particle is of the order of 
micrometers. Next, in order to realize the periodic boundary conditions 
in the direction of the external force, it might be a good 
method to place all 
the particles under a rotating optical tweezer with a velocity $v$ 
\cite{Seifert}. Indeed, by observing the system in the moving frame $v$, 
we can confirm that the spatially homogeneous force $f=\gamma v$ is produced 
with periodic boundary conditions.  (See Fig. \ref{periodic_bc}.)
We expect that such an implementation is possible due to 
the recent development of the optical technology \cite{Grier2}.

In conclusion, we have demonstrated that the fluctuation-response 
relation given in Eq. (\ref{FRR}) is plausible for  many Brownian 
particles under an external driving force  when we focus on a direction
transversal to the driving force.  If this relation is valid for
any average density, one can obtain a formula that relates the intensity
of density fluctuations in the transversal direction 
to the derivative of a chemical potential with respect to the 
density \cite{HS1}.  Then, by measuring the work required to change
the system size in the transversal direction, one may confirm
the Maxwell relation, which ensures the existence of a thermodynamic
function extended to non-equilibrium steady states \cite{SST,HS1}. 
In this manner, we will confirm Einstein's formula for the
system we study.  This provides a realistic and nontrivial
example for a framework of steady-state thermodynamics.

We thank Masaki Sano, Yoshihiro Murayama, and Kumiko Hayashi
for their helpful suggestions. This work was supported by 
a grant (No. 19540394) from the Ministry of Education, Science, 
Sports and Culture of Japan.

\end{document}